\documentclass[final,12pt]{elsarticle}

\usepackage{graphicx}
\usepackage[font=footnotesize]{subfig, caption}
\DeclareGraphicsExtensions{.pdf,.jpeg,.png,.jpg}

\usepackage{amssymb}

\usepackage{algorithm ,algorithmic}

\newcommand\coding[1]{\textbf{\texttt{#1}}}

\usepackage{listings}

\usepackage{lstlang0}
\let\origthelstnumber\thelstnumber
\makeatletter
\newcommand*\Suppressnumber{%
  \lst@AddToHook{OnNewLine}{%
    \let\thelstnumber\relax%
     \advance\c@lstnumber-\@ne\relax%
    }%
}

\newcommand*\Reactivatenumber{%
  \lst@AddToHook{OnNewLine}{%
   \let\thelstnumber\origthelstnumber%
   \advance\c@lstnumber\@ne\relax}%
}
\usepackage{lineno}

\usepackage[nolist]{acronym}

\usepackage{todonotes}
\newcommand{\tododl}[1]{\todo[backgroundcolor=pink,size=\tiny]{DL: #1}}

\usepackage{hyperref}

\usepackage{verbatim} 

\journal{Future Generation Computer Systems}

\begin{document}

\begin{frontmatter}


\title{Parallelizing Machine Learning as a Service\\for the End-User}



\author[unibo]{Daniela Loreti\corref{cor}}
\author[unimore]{Marco Lippi}
\author[unibo]{Paolo Torroni}
\address[unibo]{DISI -- University of Bologna}
\address[unimore]{DISMI -- University of Modena and Reggio Emilia}
\cortext[cor]{Corresponding author}

\begin{abstract}
As \ac{ML} applications are becoming ever more pervasive, fully-trained systems are made increasingly available to a wide public, allowing end-users to submit queries with their own data, and to efficiently retrieve results. With increasingly sophisticated such services, a new challenge is how to scale up to ever growing user bases. In this paper, we present a distributed architecture that could be exploited to parallelize a typical \ac{ML} system pipeline. We propose a case study consisting of a text mining service, and discuss how the method can be generalized to many similar applications. 
We demonstrate the significance of the computational gain boosted by the distributed architecture by way of an extensive experimental evaluation.
\end{abstract}


\end{frontmatter}


\section{Introduction}\label{sec:introduction}

In the last decade, \ac{ML} has undoubtedly become one of the hottest topics in computer science. By exploiting big data collections, \ac{ML} algorithms are now being implemented and deployed on a large scale across countless application domains, including health-care, transportation, speech analysis, computer vision, market analysis, life sciences, and many others~\cite{sejnowski2018deep}.

Recently, \ac{ML} applications have been moving to the cloud, in order to exploit high performance parallel and distributed computing, which has given rise to the concept of \ac{MLaaS}~\cite{ribeiro2015mlaas}. This usually refers to the availability of online platforms and frameworks, which has enabled to implement in the cloud all the customary stages of a \ac{ML} pipeline. Such stages include, for instance, input pre-processing, feature extraction, and in particular the training phase, which is usually the most expensive from a computational and  memory consumption viewpoint~\cite{yao2017complexity}.

In this paper, we turn our attention to a particular aspect of \ac{MLaaS}, that of deploying and parallelizing  systems that have already been trained, and need to be made available to (possibly many) end-users. Although not computationally expensive as the training phase, prediction and forecasting tasks may be nonetheless burdensome in terms of resources, especially when results must be delivered within a close deadline, or when the service has to be made available to a wide public.
In this setting, an infrastructure for \ac{MLaaS} should support large-scale distributed batch processing, as well as run-time stream processing. Besides, scalability and fault tolerance are fundamental requirements. In this regard, since its first formulation in 2004, the MapReduce \cite{DBLP:conf/osdi/DeanG04} distributed programming model has gained significant diffusion in the big data research community. This success is mainly due to its simple yet powerful and intrinsically parallelizable paradigm. Several distributed computing engines have been proposed since then, to support the development of distributed programs for batch processing of very large data collections, providing autonomous fault-tolerant mechanisms and run-time infrastructure scaling capabilities. The latest evolution of these frameworks~\cite{storm, flink, spark} offers stream processing support, alongside with the more traditional batch processing, and leverages different programming paradigms, in addition to MapReduce. All that together greatly simplifies the implementation of efficient distributed applications for the analysis of big data flows.

Our aim with this paper is to show how a MapReduce-inspired programming paradigm can be used to improve the performance and scalability of the \ac{ML} pipeline, by parallelizing the customary steps of the prediction process and supporting the development of ready-to-use pre-trained services for the end user. These are our main contributions:
\begin{itemize}
    \item A structural characterization of systems addressing the prediction task of different \ac{ML} applications, from a MapReduce-inspired parallelization viewpoint.
    \item The  detailed discussion of a concrete application of the outlined structure in the natural language processing domain, specifically in the area of argumentation mining, together with an empirical evaluation of the performance and scalability of the approach.
    \item A discussion of the issues and challenges that must be addressed when parallelizing the identified general \ac{ML} pipeline, as well as the features that might induce significant enhancements in performance and desirability of the offered \ac{ML} service.
   
\end{itemize}

The paper is structured as follows. Section~\ref{sec:related} discusses related work, introducing the concept of \ac{MLaaS} and its interpretations in various domains of computer science. Section~\ref{sec:prediction} describes scenarios where the proposed parallel architecture could play a role. Section~\ref{sec:argumentation} focuses on the proposal of \ac{MLaaS} for the end user by presenting a case study in the area of argumentation mining. Section~\ref{sec:parallel}  illustrates the parallel architecture. Section~\ref{sec:experiments} presents its empirical evaluation, whereas Section~\ref{sec:discuss} discusses the challenges  the argumentation mining case study has helped bringing up. Section~\ref{sec:conclusions} concludes.

\section{Related Work}\label{sec:related}

\ac{MLaaS} is a phrase found in various areas of computer science, where it is used to refer to various concepts.
A significant body of literature focuses on the description and analysis of platforms that implement a whole \ac{ML} pipeline. That may include, for instance, the capability to perform data pre-processing and feature selection, to choose the best-performing classifier, to train a model, and to predict the outcomes on query data. For example, Yao et al.~\cite{yao2017complexity} compare the performance and complexity of several solutions for building \ac{MLaaS} applications implementing the entire \ac{ML} pipeline. Similarly, Chan et al.~\cite{li2017scaling} describe the distributed architecture exploited in \ac{ML} applications in Uber, with a focus on model training and features selection. 
A complete architecture for \ac{MLaaS} is also described by Ribeiro et al.~\cite{ribeiro2015mlaas} who present a specific analysis on three \ac{ML} classifiers (multi-layer perceptrons, support vector machines, $K$-nearest neighbors). In each of these works, however, the parallelization of the final prediction stage is only superficially addressed, or ignored altogether, whereas we believe that a \ac{ML} tool provided as-a-service to a wide public of end-users cannot disregard the parallelization of this last step (although it is generally less computationally expensive then the previous ones). In particular, we argue that the prediction stages of different ML applications have common characteristics, which make the adoption of a MapReduce-oriented approach \cite{DBLP:journals/cacm/DeanG08} particularly suitable for the purpose.
The need for a \ac{ML} tool as a service through the adoption of MapReduce is also envisaged by other scholars. For instance, Chan et al.~\cite{DBLP:conf/cikm/ChanSSC13} and Baldominos et al.~\cite{DBLP:conf/cibd/BaldominosASI14} do so while focusing on the features that such a tool should expose, rather than on techniques to obtain scalability.
The application of MapReduce to the \ac{ML} pipeline comes with the great advantages brought by distributed computing architectures, which allow the developer to focus on the implementation of its parallel program while disregarding lower-level architectural and infrastructural details, such as the coordination between nodes, the employment of heterogeneous hardware in the same data-center \cite{DBLP:conf/osdi/ZahariaKJKS08,DBLP:journals/tpds/ChengRGJZ17} or even the use of multiple cloud data-centers \cite{DBLP:journals/ijcc/AntoniuCBDFGPSTBTBBCKNS13,DBLP:conf/ucc/Clemente-Castello15,DBLP:conf/ucc/LoretiC15,DBLP:conf/hpcc/LoretiC15}.

Another line of research that uses the \ac{MLaaS} term, focuses instead on the parallelization of training algorithms for \ac{ML} systems.
For example, a multicore implementation for the training of many \ac{ML} systems, which exploits the MapReduce paradigm, is presented by Chu et al.~\cite{chu2007map}.
Sergeev et al.~\cite{DBLP:journals/corr/abs-1802-05799} present an interesting framework to enable faster and easier distributed training in TensorFlow \cite{tensor}.
Tamano et al.~\cite{tamano2011optimizing} illustrate an approach to job scheduling optimization for MapReduce tasks in \ac{ML} applications.
The challenges and opportunities of exploiting \ac{MLaaS} in the context of the Internet of Things are discussed by Assem et al.~\cite{assem2016machine}, again with a focus on the aspects related to training and classifier selection.

Fewer strands of work are instead dedicated to the performance of parallel algorithms for already trained and deployed \ac{ML} models.
\citeauthor{xu2015making}~\cite{xu2015making} propose a software architecture that encompasses model deployment for real-time analytics. Their emphasis is on the processing of big data collections, including RESTful web services for data wrapping and integration, dynamic model training, and up-to-date prediction and forecasting. The resulting architecture relays on Spark's MLlib library to offer a set of already-parallelized, popular \ac{ML} functions to the end-user. Given the particular focus of their study, \citeauthor{xu2015making} do not propose a general methodology for distributing the \ac{ML} pipeline, and in particular for parallelizing the prediction task of possibly more sophisticated applications and leave it to the developer to address such a problem. Similarly, \citeauthor{harnie2017scaling}~\cite{harnie2017scaling} use the Apache Spark technology \cite{spark} to achieve the desired scalability in chemoinformatics applications, which is also a choice we made in our work. However, as their main target is obtaining performance enhancements through the parallelization of the scientific application at hand, they do not provide a general methodology for parallelizing prediction services. Nonetheless, it can be observed how the software implementation proposed by \citeauthor{harnie2017scaling} (composed of two initial mapping tasks and a subsequent aggregation by reduction) can be easily framed in the methodology we propose (see Section~\ref{sec:prediction}).
The focus of other works such as that by Hanzlik et al.~\cite{hanzlik2018mlcapsule} is on model deployment, and in particular on problems related to model stealing and reverse engineering. In the envisaged scenario, the ML model is made available to the client for offline prediction on sensitive data that cannot leave the client's data center. The authors devote great attention to the performance of the security-enhanced forecasting step, but do not address the parallelization of such a task.

The problem of parallelizing sophisticated artificial intelligence-based reasoning engines has been studied by Loreti et al.~\cite{DBLP:conf/wosp/LoretiCCM17,DBLP:journals/fgcs/LoretiCCM18} in the domain of business process compliance monitoring. However, the focus of such study lies in the input data-set partitioning strategies, whereas the present study aims to identify common patterns in various \ac{ML} tasks, which could lead back to a MapReduce-inspired approach.

\section{Prediction as a Service}\label{sec:prediction}

\begin{figure}[!b]
\centering
\includegraphics[width=0.9\columnwidth]{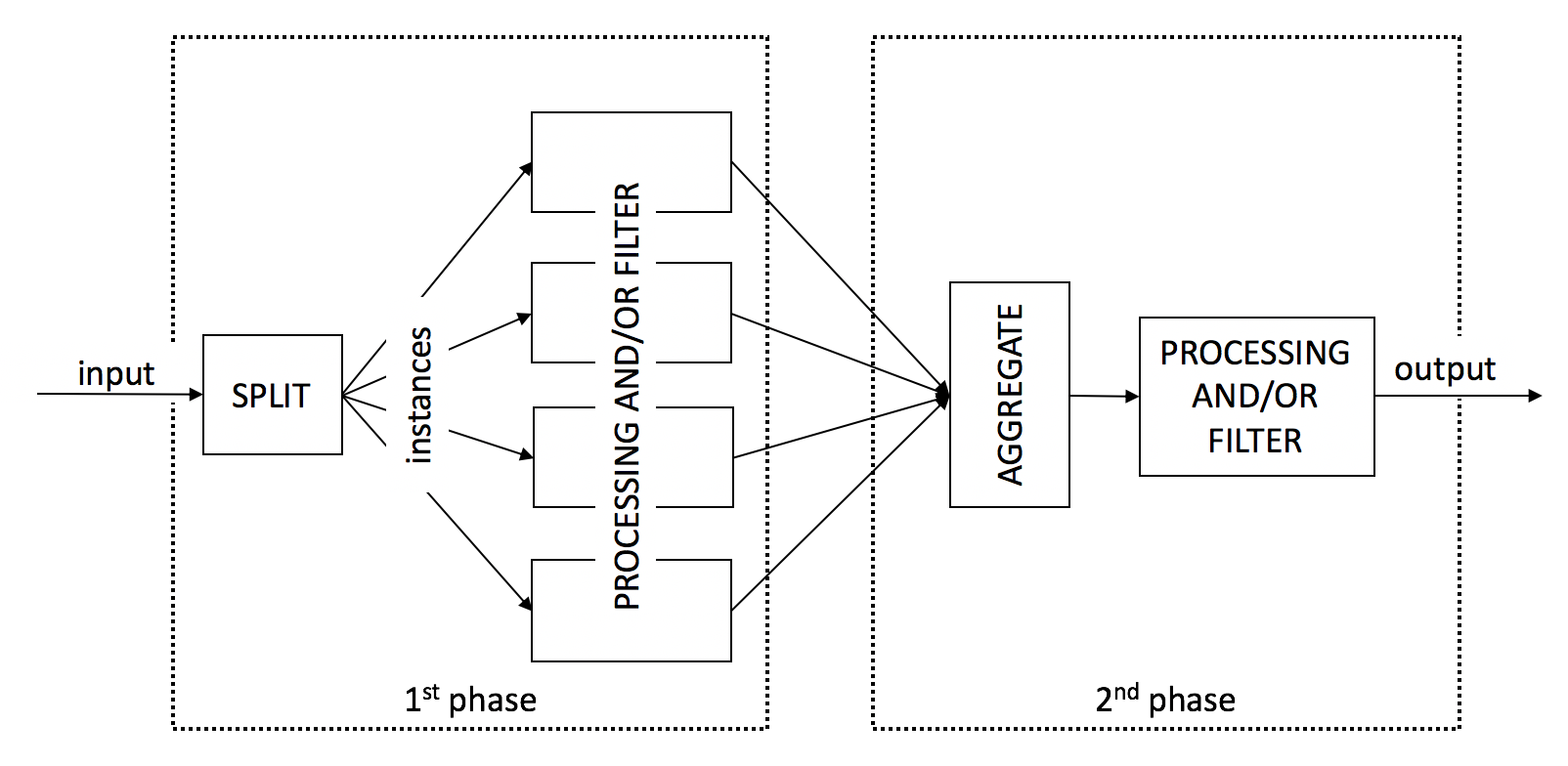}
\caption{Generic pipeline of steps of a \ac{ML} application}
\label{fig:g-pipeline}
\end{figure}
The typical pipeline of a \ac{ML} system  to parallelize and distribute has the following characteristics.
A collection of data is given as input, either as a set of batches, or as a continuous stream. As Figure~\ref{fig:g-pipeline} illustrates, a first stage of computation is present, where all the instances of the input dataset need independent processing. Such independence enables the distribution of the computation load across many nodes. This set up is common to many application domains.
For instance, a great deal of natural language processing tasks consider independent input elements such as single sentences, paragraphs, or documents. Among such tasks we shall mention sentiment analysis, whose goal is to assign a \textit{sentiment} to a given segment of text, being it a tweet, a post on a social network, or a comment to a newspaper article. Document categorization aims to \textit{classify} a piece of text into one or more semantic categories. Fake news detection, as well as many other tasks, aim to \textit{detect} sentences or text portions with certain characteristics~\cite{manning1999foundations}.
The same can be said of many important tasks in computer vision. For example, the goal of  image tagging or image classification  is to assign a label (or a set of labels) to a given image. In video surveillance~\cite{steger2018machine}, as well as in many other video processing tasks, the computation is often carried out at the level of single frames, or small batches of frames.
In other domains, such as bioinformatics, chemoinformatics, or genomics data analysis, several different predictors can be applied to input data, such as protein or DNA sequences, so as to classify different properties of single sequence elements~\cite{baldi2001bioinformatics}.

The output of this first stage of the pipeline if often used as input to a second stage. In general, the first stage could be considered a sort of filtering, or detection phase, whereas the second stage works as a sort of aggregation phase, where the instances detected during the first round are matched and compared with one another. 
\begin{figure*}[t!]
\centering
\subfloat[]{\includegraphics[width=.8\textwidth]{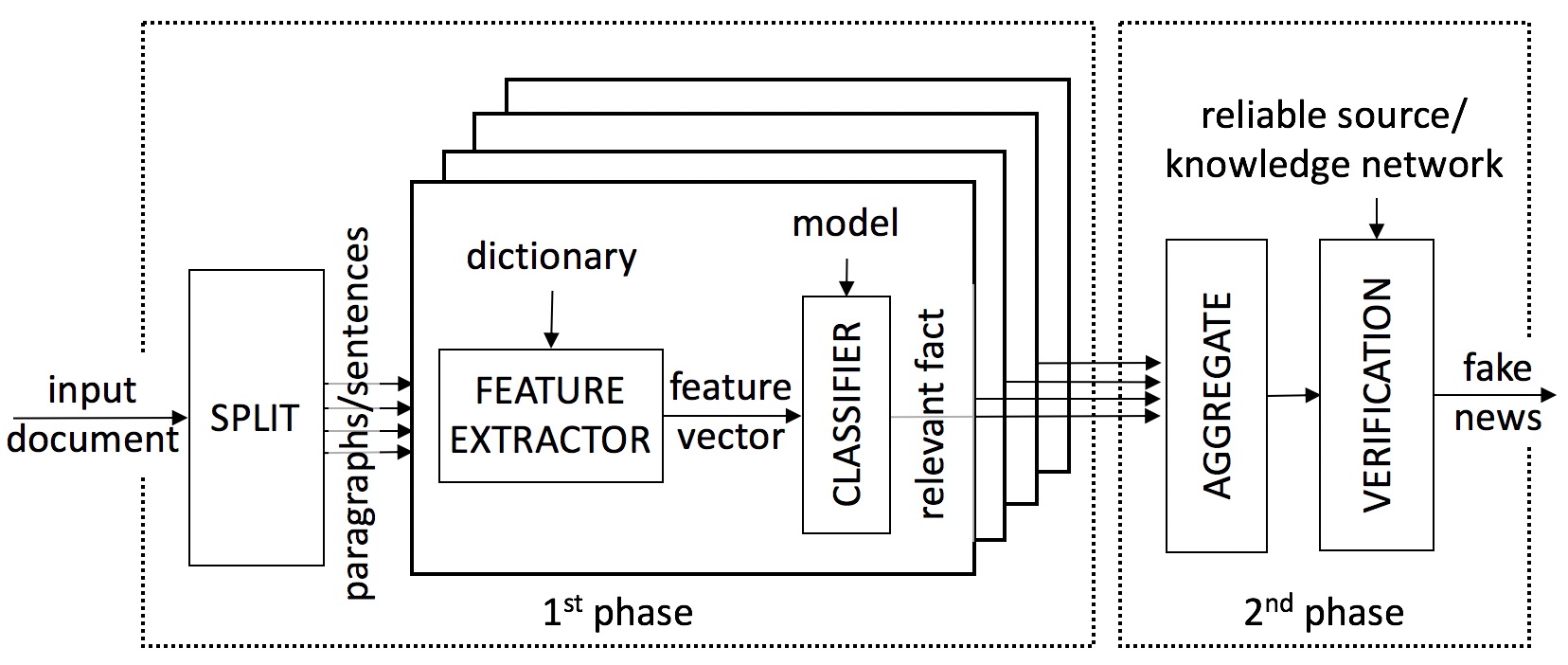}
\label{fig:fake}}
\hfill
\subfloat[]{\includegraphics[width=.8\textwidth]{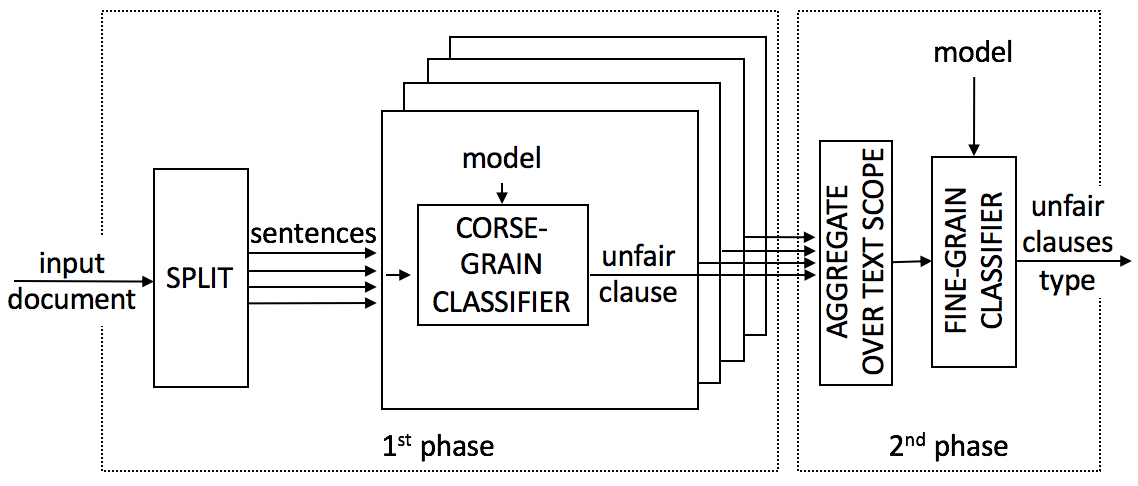}
\label{fig:unfair}}
\hfill
\subfloat[]{\includegraphics[width=.8\textwidth]{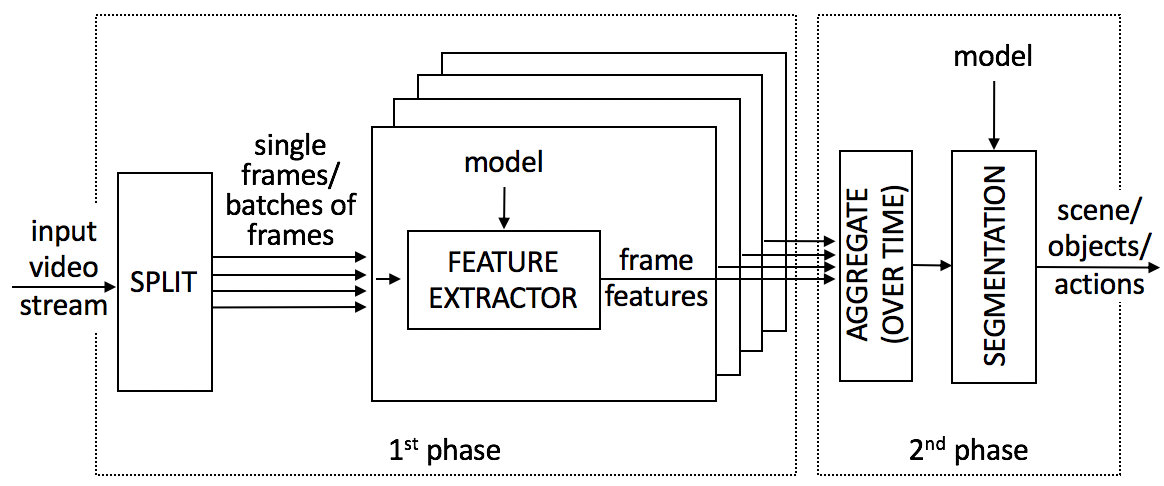}
\label{fig:video}}
\caption{Examples of pre-trained \ac{ML} applications: detection of fake news (Fig. \ref{fig:fake}), unfair clauses in legal documents (Fig. \ref{fig:unfair}), and video segmentation. The steps of these pipelines recall the MapReduce programming paradigm.}
\label{fig:pipiline-examples}
\end{figure*} 
That is the case in some natural language processing setups, where the text segments selected during the first phase are processed by a clustering algorithm, such as topic modeling, or by a further categorization stage. For instance, single sentences could be first classified according to coarse-grained categories, and then classification could be further developed into fine-grained categories. That happens with fact-checking systems, where candidate fake news items are first selected during an initial processing phase, and then a fact-checking algorithm is applied to a collection of instances found in the first phase~\cite{Ciampaglia15}, possibly relaying on a trustworthy source. Fig \ref{fig:fake} sketches the steps of this pipeline.
In some consumer-oriented applications~\cite{Lippi2019nature}, such as the detection of unfair contract clauses~\cite{Lippi2019claudette}, a first stage could identify sentences expressing potentially unfair clauses through a coarse-grained classifier (see Figure~\ref{fig:unfair}), whereas a second stage could additionally recognize the category of unfairness among a set of pre-defined cases (e.g., arbitration, limitation of liability, etc.), usually on the basis of the context provided by other relevant sentences within a certain document scope. 
In bioinformatics, the output of the predictors developed during the first stage are often aggregated into a higher-level set of predictions, that combine the information coming from the different classifiers~\cite{harnie2017scaling}. In computer vision, video segmentation is typically performed on frame groups, resulting from a first processing stage~\cite{tekalp2015digital}. Typically, as shown in Fig. \ref{fig:video}, the frames are initially analyzed to extract low-level features such as brightness, contrast, etc. and later aggregated over a temporal basis to extract higher-level information (e.g., scenes, objects, actions, etc.).

The general pipeline described so far recalls the principles of the MapReduce programming paradigm \cite{DBLP:journals/cacm/DeanG08}.
MapReduce is a well-known technique for simplifying the parallel execution of software, whereby the input data is partitioned into an arbitrary number of slices, each exclusively processed by a \emph{mapper} task emitting intermediate results in the form of key/value pairs. The pairs are then passed on to other tasks, called \emph{reducers}, which are in charge of merging together the values associated with the same key, and emitting the final result. 
As Figure~\ref{fig:g-pipeline} illustrates, the first phase of the general \ac{ML} pipeline could indeed be modeled as a mapping task, where each instance undergoes the same processing to emit a key/value pair. The corresponding reduction task is then carried out in the second phase of the pipeline, where the outputs of the first phase are merged and processed together.

Programs reformulated according to the MapReduce model can be automatically parallelized by means of a big data analytics platform~\cite{Singh2014}, which can turn a collection of computing nodes into a distributed infrastructure able to automatically spread (and balance) the execution tasks across the data center. Most of these platforms supply mechanisms to scale up or down the cluster on demand and perform fault detection and recovery from failures at runtime. All these features seem particularly beneficial to a \ac{ML} tool provided as-a-service to a wide public.

In an effort to ground the general discussion and illustrate in concrete how \ac{MLaaS} can be enabled by a MapReduce-oriented approach, the section that follows is devoted to a particular case study. The case study is a natural language processing application in the area of argumentation mining~\cite{DBLP:journals/toit/LippiT16}, consisting of an argument component identification task followed by an argument structure prediction task, whose goal is to identify links and relations between argument components.

\section{A Case Study}\label{sec:argumentation}
Argumentation mining~\cite{DBLP:journals/toit/LippiT16} defines the task of extracting arguments from unstructured text by automated analysis methods. Such a task is usually, but not necessarily, referred to a specific genre, such as legal texts, persuasive essays, scientific literature, etc. One notable and increasingly popular application area where argumentation mining plays a key role is that of debating technologies \cite{mirkin-etal-2018-listening}, where the data to be processed may consist of ``static'' Wikipedia pages as well as streamed audio signals. Owing also to the variety of addressed genres, the argumentation mining literature offers several alternative definitions of argument~\cite{Peldszus2013}, with varying degrees of sophistication. For the aims of this work, and without loss of generality, we shall refer to a common and rather basic claim/premise argument model~\cite{walton1990reasoning}, whereby an argument is composed by an assertion, or statement (the \textit{claim}) supported by one or more \textit{premises}, and the inference that connects claim and premises.

\begin{figure}[!t]
\centering
\includegraphics[width=0.99\columnwidth]{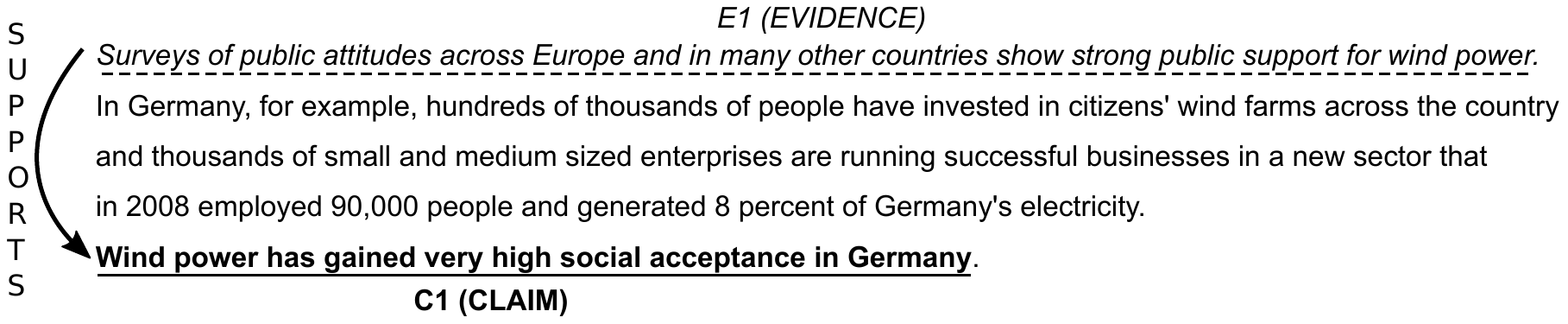}
\caption{Example of the output of an argumentation mining system, taken from the IBM Wikipedia corpus~\cite{rinott2015show}. Given a text, this system detects evidence (in italics), claims (in bold), and their support links.}
\label{fig:am-example}
\end{figure}

Argumentation mining includes several sub-tasks, spanning from claim and evidence detection (i.e., the identification of argument components), to attribution (i.e., detection of the authorship of an argument), to increasingly challenging tasks such as the prediction of the relations between arguments (\textit{argument structure prediction}), the inference of implicit argument components (\textit{enthymemes}), and so forth. Figure~\ref{fig:am-example} offers an example of a possible output produced by an argumentation mining system.

One such system is \ac{MARGOT} \cite{DBLP:journals/eswa/0001T16}. \ac{MARGOT} exploits a combination of \ac{ML} and natural language processing techniques in order to perform argumentation mining on unstructured texts of various genres. In particular, \ac{MARGOT} exploits a \ac{SVM}-based method for context-independent claim detection \cite{DBLP:conf/ijcai/LippiT15} using tree kernels, and extends its application to evidence detection, and to the identification of their boundaries.

The initial version of \ac{MARGOT} follows a pipeline of subsequent stages, by initially segmenting the text into sentences, then performing the detection of argumentative sentences, that is sentences that include argument components (claims or evidence), and finally identifying the boundaries of each argument component. The current  \ac{MARGOT}  prototype \cite{margot}, also available as a web server,\footnote{See \url{http://margot.disi.unibo.it}} adopts the ``traditional'', sequential architecture. It analyzes each input sentence individually, thereby processing the document as a whole in a sequential manner, in order for each sentence to undergo all the subsequent steps of the pipeline. It does not address argument structure prediction. It is worthwhile noticing that \ac{MARGOT}'s training phase is executed offline, once and for all. It results in two trained models, one for claim detection and one for evidence detection, which are then deployed for use of the online server.

In this work, taking \ac{MARGOT} as a case study, we present a more advanced version of the system, where the claim and evidence detection stage is followed by the application of a further \ac{SVM}-based step, aimed to detect the existing support links between all possible pairs of claims and evidence found in each document. The offline training of this additional stage resulted in a third model, which was deployed for our experimental analysis.

\begin{figure}[!b]
\centering
\includegraphics[width=\columnwidth]{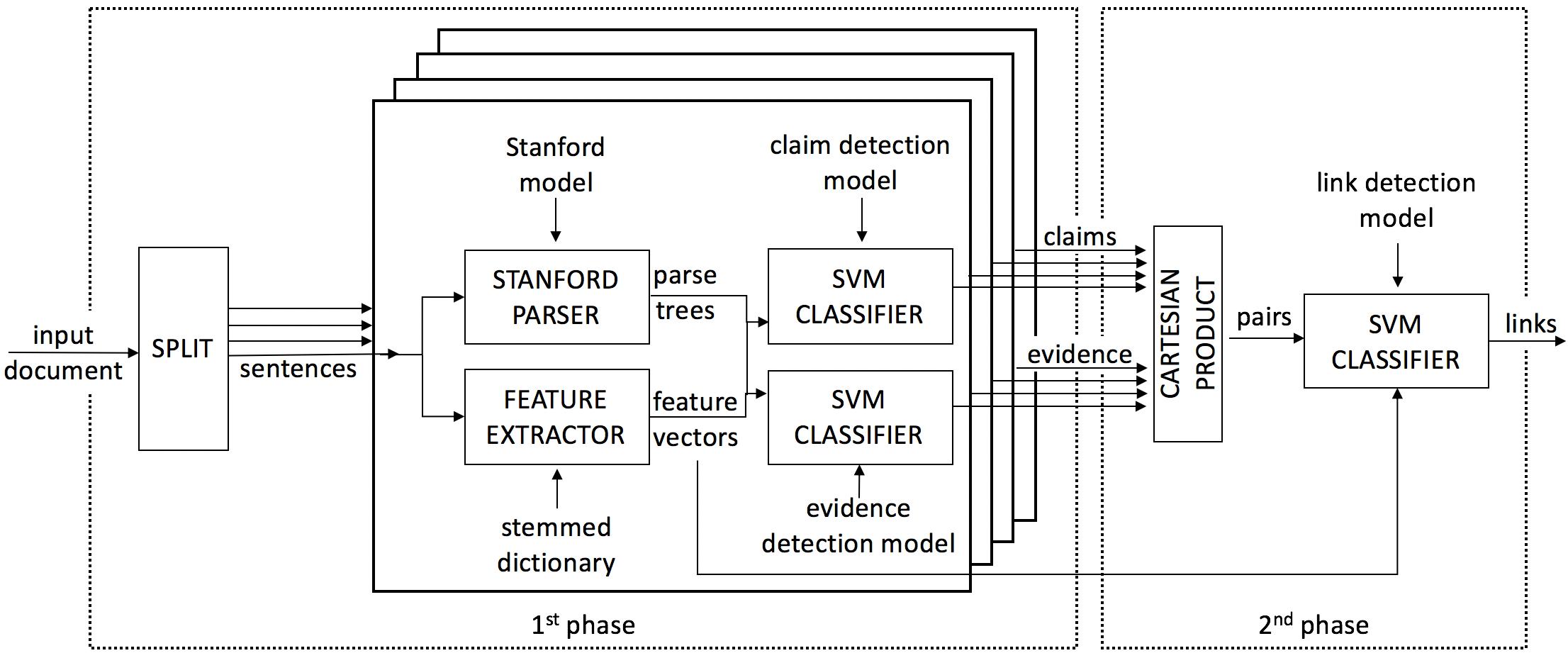}
\caption{The pipeline of steps implemented in \ac{MARGOT}}
\label{fig:pipeline}
\end{figure}

Following the model introduced in Section~\ref{sec:prediction}, we describe \ac{MARGOT}'s pipeline as composed of two subsequent phases. A first phase, shown in Figure~\ref{fig:pipeline}, processes each sentence of the input file using the Stanford Parser~\cite{DBLP:conf/acl/ManningSBFBM14}, a largely successful third-party software package. \ac{MARGOT} uses the Stanford Parser to obtain a first set of features, that is, the trees encoding the grammatical structure of a sentence, known in the literature as \textit{constituency parse trees}. From the same sentence, \ac{MARGOT} also extracts the bag-of-words feature vector, which represents a binary encoding of its words. The two sets of features are passed on to two different classifiers employing the \ac{SSTK}. The aim of these classifiers is to identify claims and evidence, respectively. All and only the sentences containing the identified claims/evidence are then sent to the second phase of the pipeline, which considers all the possible \textit{(claim, evidence)} pairs in each file and calls another \ac{SVM}-based classifier to detect the possible links between the two (link detection).

Given the complexity of the analysis to be carried out, especially when large files are mined, the sequential execution of \ac{MARGOT}'s pipeline can be highly time- and resource-consuming. The problem is exacerbated when the number of arguments detected in the first phase is large, since this entails to consider, for each input file, the Cartesian product of large sets of the detected premises and claims. Furthermore, as we envisage the future necessity of argumentation mining analysis as a service, it is likely that such a service would be required to consider streams of input text instead of documents already materialized in a certain disk location. The run-time nature of stream processing further amplifies the need for scalable and reliable architectures to support argumentation mining as a service.

Nonetheless, as prescribed by the general model in Figure~\ref{fig:g-pipeline}, the parser and the feature extractor of the first phase can process each sentence independently from the others, whereas the Cartesian product in the second phase operates as a pair-wise sentence aggregator.
This observation suggests a way to distribute the computational load of \ac{MARGOT}'s pipeline on a network of computing nodes, leveraging a MapReduce-oriented approach and an engine for large-scale batch and stream processing.

\section{A Parallel Architecture}\label{sec:parallel}

Among the existing variety of MapReduce-oriented engines for large-scale data processing, some offer the possibility to analyze batches of documents already materialized in a certain location \cite{hadoop}; others only deal with the processing of data flows \cite{storm,samza,dataflow}; whereas a restricted number offer the possibility -- particularly desirable for the current work -- to operate on both batches and streams \cite{flink, spark}. 
For the purposes of this work, and without loss of generality, we will describe how \ac{MARGOT} can be re-implemented to be automatically executed on a distributed infrastructure using the facilities provided by Apache Spark~\cite{spark}.
Apache Spark has become increasingly popular in the last years also because it allows a developer to write her application in several different languages, without forcing her to think in terms of only map and reduce operators. Its good performance~\cite{DBLP:conf/bigdataconf/VeigaEPTT16,DBLP:conf/ipps/ChintapalliDEFG16,DBLP:conf/iccS/SamosirIH16} and resilience to faults~\cite{DBLP:conf/globecom/LopezLD16} has been empirically verified by various studies.
We will first consider the case of large batches of input documents, and later refine the algorithm in order to accommodate streams as well.

\subsection{\acs{MARGOT} for batch processing}
\label{subsec:batch}
In case of batch processing, all the documents to be analyzed are already present in a certain (centralized or distributed) disk location, and the data analytics infrastructure is in charge of spreading (and balancing) the computation load on the available nodes. 

\begin{figure}[b!]
\centering
\includegraphics[width=\columnwidth]{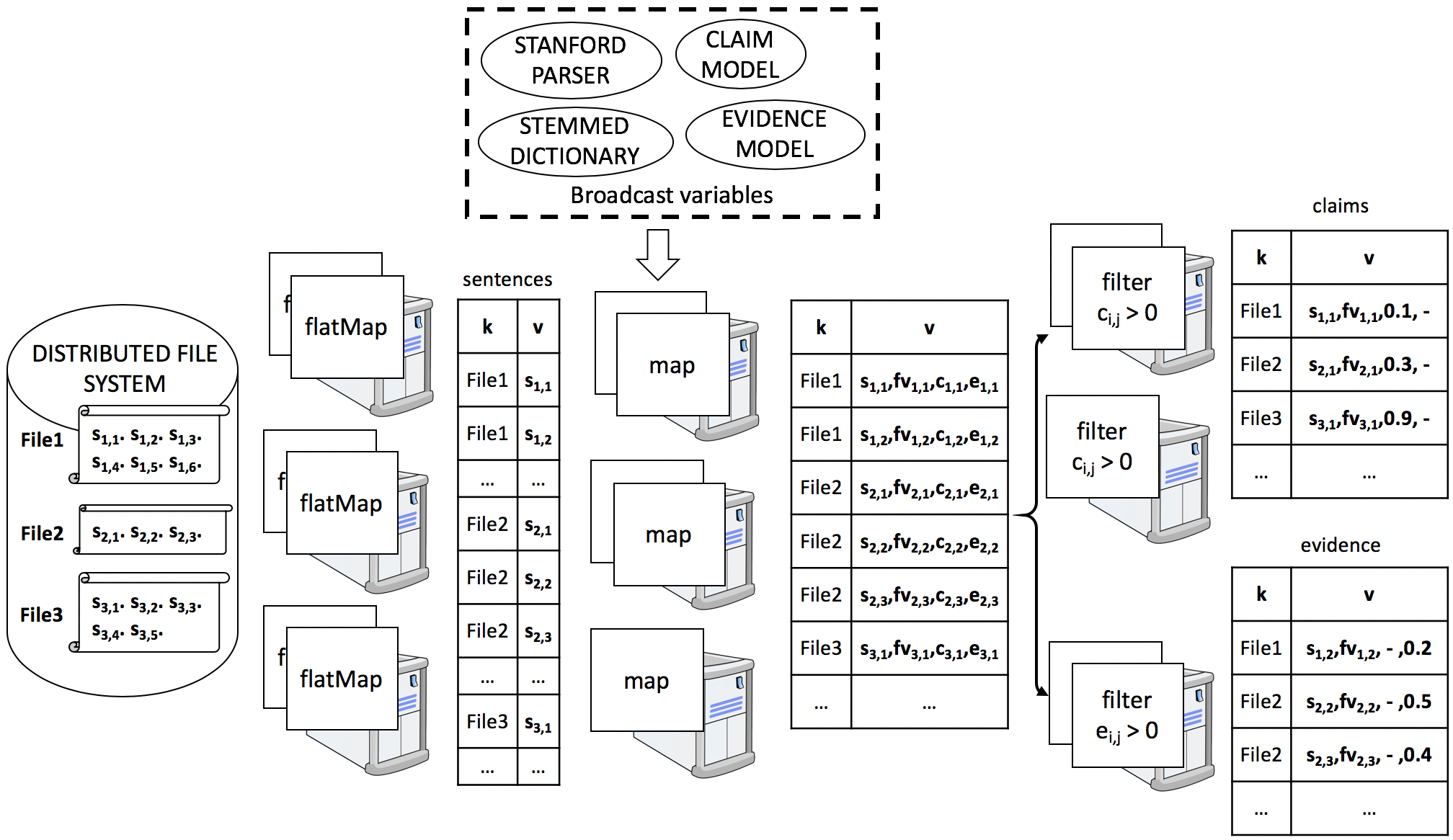}
\caption{Distributed version of the first phase of \ac{MARGOT}, working on batch input.}
\label{fig:1st-phase}
\end{figure}

As Figure~\ref{fig:1st-phase} illustrates, in the first phase of \ac{MARGOT}'s pipeline the files to be analyzed are split into sentences. A collection of ($key,value$) pairs is thus produced, where the key is the file name and the value is the sentence found. 

Then, the core operations of the first phase are performed on each sentence independently by applying a map function. This operation extracts the parse tree and the bag-of-words feature vector, and passes them in input to two third-party classifiers that emit, for each sentence, a claim and an evidence score, respectively. As these functions require sizable models (the parsing model, the stemmed dictionary and the claim/evidence models) to operate on each sentence, we apply the general suggestion of loading such models once at the beginning of the computation. The objects produced are sent to all the computing nodes, by leveraging the concept of \textit{broadcast variable} offered by Apache Spark i.e., an immutable shared variable which is cached on each worker node of the Spark cluster.
The output of the map function is again a ($key,value$) pair with the same key (the file name), but with the values (which hosted just the sentence in the input) now enriched by the phrase feature vector, claim and evidence score.

Finally, two filters are applied to select only claims and premises. Indeed, two different collections of pairs are produced: one containing elements with positive claim score, and another with positive evidence score. 

The details of the first phase implementation on a distributed MapReduce-oriented platform are presented in Listing~\ref{lst:1st-phase} following an Apache Spark-inspired approach with lambda functions. 
It is worthwhile underlining that, since the \ac{SVM}-based classifiers are realized by third-party C software \cite{DBLP:conf/ecml/Moschitti06} that cannot be directly converted into a Spark broadcast variable, we attempt to minimize the initial overhead of loading these external software by resorting to a \coding{mapPartitions} function. 
Differently from \coding{map} and \coding{mapValues} functions, which are executed for each sentence (as in line \ref{lst:1st-phase:mapValues}), \coding{mapPartitions} operates on a collection of sentences, which is a partition of the whole sentences in the input files. The size of this partition is optimized by the underlying infrastructure, based on the number of  available cores. As a consequence, the call to external \ac{SVM}-based classifiers and the loading of the models in lines \ref{lst:1st-phase:loadC} and \ref{lst:1st-phase:loadE} is not repeated for each sentence (which would be highly inefficient) but for each group of sentences the infrastructure has partitioned the input in.
We shall remark that this solution could be applied to other \ac{MLaaS} applications too, in order to deal with the frequent challenge of reducing the overhead of third-party software invocation. 

Since the result of \coding{mapPartitions} is later accessed by two different filters, cluster-wide caching (line~\ref{lst:1st-phase:cache}) is employed to avoid a duplicated computation of the same \coding{mapPartitions} stage. 

\Suppressnumber
\begin{lstlisting}[language=Scala,label={lst:1st-phase},caption={Distributed implementation of \ac{MARGOT}'s first phase (batch mode).}, 
float=tp, floatplacement=tbp,
columns=flexible,breaklines=true,breakatwhitespace=true]
Input: inputDir /*directory containing the input documents*/, stanfordModel /*model for the Stanford Parser*/, stemmedDict /*dictionary of stems for feature extraction*/, claimModel, evidenceModel /*SVM classification models*/
Output: claims, evidence /*sentences containing claims/evidences*/  |\Reactivatenumber|
stanfordParser = LexicalizedParser.loadModel(stanfordModel)
sparkContext.broadcast(stanfordParser)
sparkContext.broadcast(stemmedDict)
sparkContext.broadcast(claimModel)
sparkContext.broadcast(evidenceModel)
phrases = sparkContext
  .wholeTextFiles(inputDir) //emit (fileName,fileContent) pairs
  .flatMapValues(fileContent => 
    fileContent.split(|``|[.!?]|''|)) //split into sentences and emit (fileName,sentence) pairs
  .mapValues(sentence => { |\label{lst:1st-phase:mapValues}|
    tree = stanfordParser.apply(sentence) // generate parsing tree 
    fv = FeatureExtractor
        .createBagOfWords(tree,dmStemmed) // generate feature vector
    (sentence, tree, fv) })
  .mapPartitions((partition) => {
    svmC = SvmClassify.load(claimModel) |\label{lst:1st-phase:loadC}|
    svmE = SvmClassify.load(evidenceModel)|\label{lst:1st-phase:loadE}|
    outList = List()
    l = partition.toList 
    l.foreach( x => { //compute claim and evidence score of each sentence in partition
      filename = x._1
      sentence = x._2._1
      tree = x._2._2
      fv = x._2._3
      claimScore = svmC.getScore(tree, fv)
      evidScore = svmE.getScore(tree, fv)
      outList=(filename,(sentence,fv,claimScore,evidScore))::outList  })
    outList.iterator })  //emit a partition with the elements of the novel list
  .cache |\label{lst:1st-phase:cache}|
claims = phrases.filter(_._2._3>0) //only sentences with claimScore > 0
evidence = phrases.filter(_._2._4>0) //only sentences with evidScore > 0
\end{lstlisting}

\begin{figure}[h!]
\centering
\includegraphics[width=\columnwidth]{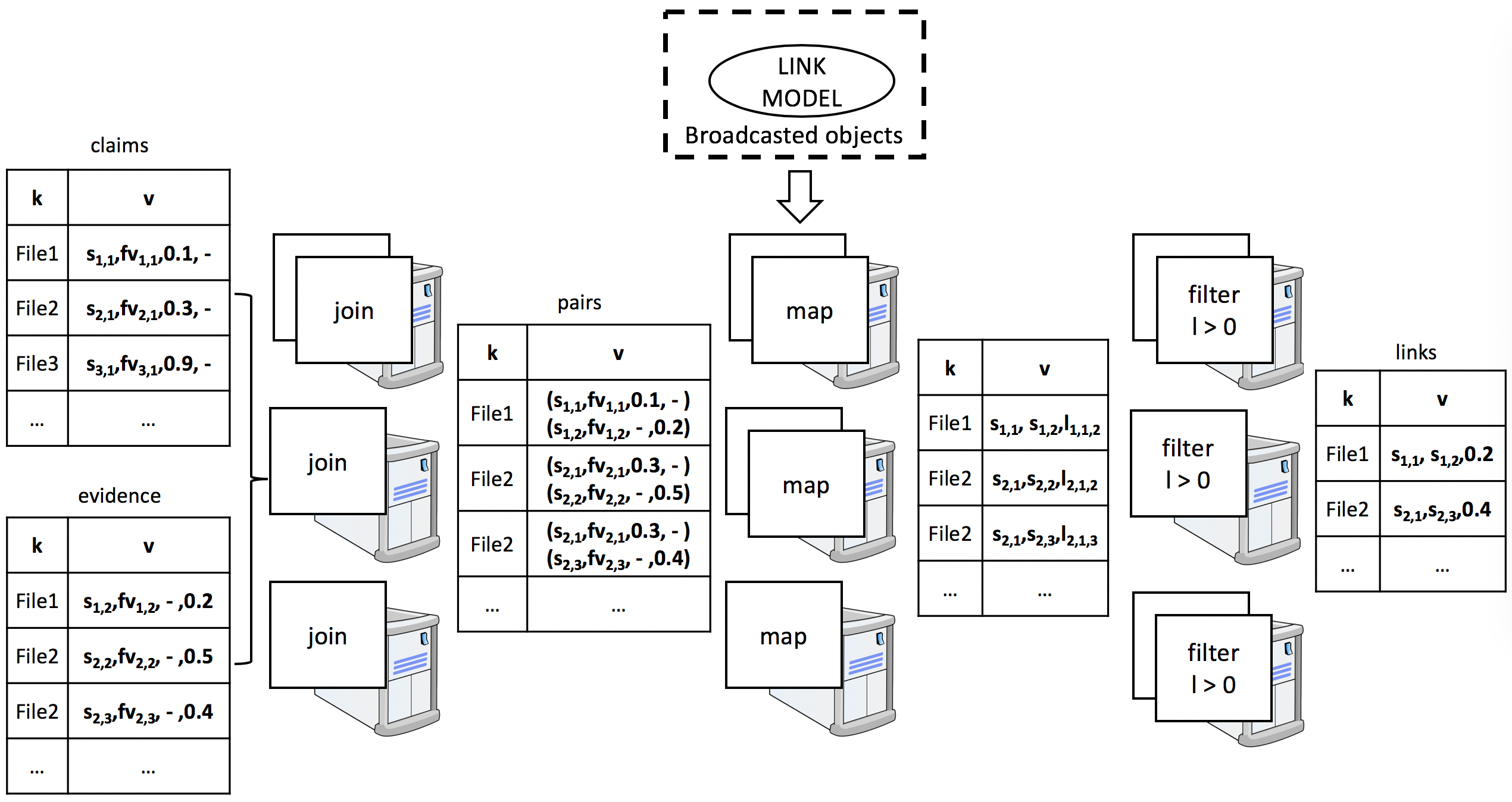}
\caption{Distributed version of the second phase of MARGOT, working on batch input.}
\label{fig:2nd-phase}
\end{figure}

The collections computed during the first phase are then merged together in the second phase of \ac{MARGOT}, as shown in Figure~\ref{fig:2nd-phase}. 
In particular, a distributed \coding{join} operation is performed, to obtain a collection of all the possible (\emph{claim,evidence}) pairs in each file. The subsequent map function considers each pair individually. Indeed, parallelizing the steps following the aggregation -- whenever possible -- should help to speed up the  computation. 
Inside each map function, the link model and the pair of feature vectors in each record are employed by a third-party \ac{SVM}-based classifier to predict a link score, indicating whether the claim and evidence in each pair are linked. As reported in Listing~\ref{lst:2nd-phase}, similarly to claim/evidence detection, a \coding{mapPartitions} function  (line \ref{lst:2nd-phase-mapPartitions}) is actually performed at this stage, aiming to minimize the initial overhead of loading the external software.
Finally, only elements with positive link scores are maintained (line~\ref{lst:2nd-phase-filter} of Listing~\ref{lst:2nd-phase}), as these represent the final output of the argumentation mining algorithm.

\begin{lstlisting}[language=Scala,label={lst:2nd-phase},caption={Distributed implementation of \ac{MARGOT}'s second phase (batch mode).},
float=tp,floatplacement=tbp,
columns=flexible,breaklines=true,breakatwhitespace=true]
Input: claims, evidence /*sentences containing claims/evidences*/, linkModel /*SVM classification model*/
Output: links /* (claim,evidence) pairs containing an argumentative link*/ 
|\Reactivatenumber|
sparkContext.broadcast(linkModel)
pairs = claims.join(evidence)
links = pairs
  .mapPartitions((partition) => { |\label{lst:2nd-phase-mapPartitions}|
    svmL = SvmClassify.load(linkModel)
    outList = List()
    l = partition.toList 
    l.foreach( x => { //compute link score of each sentence in partition
      filename = x._1
      claim = x._2._1._1
      evid = x._2._2._1
      fvC = x._2._1._2
      fvE = x._2._2._2
      linkScore = svmC.getScore(fvC, fvE)
      outList = (filename, (claim, evid, linkScore)) :: outList  })
  outList.iterator })
 .filter(_._2._3>0) //only pairs with linkScore > 0 |\label{lst:2nd-phase-filter}|
\end{lstlisting}

\subsection{\acs{MARGOT} for stream processing}
\label{subsec:stream}

The distributed algorithm presented in the previous subsections can accommodate streams of input text after only minor modifications. In particular, those operations conducted on each sentence independently, such as map and filter, can be performed on streams and batches alike, whereas the semantics of functions that merge together different collections of key-value pairs, such as the join operation in the second phase of the pipeline, needs further refining for stream processing.

Link detection on a batch of documents implicitly entails a natural definition of ``scope'' which corresponds with the document at hand. Input files should thus be considered independently, so as to identify the claims and evidence therein, enabling link detection on a per-file basis, as shown in the previous section. When dealing with a stream of data instead, two different semantics for link detection are possible.

\begin{itemize}
    \item \textit{Scope file} pairing: 
    if the stream itself contains the indication of the input file transmitted, we might be asked to pair claims with evidence in each file, with a  semantics similar to the one adopted for batch processing, thus keeping track of the currently identified claims/evidence in each file along the stream. 
    \item \textit{Scope window} pairing: if there is no explicit or implicit concept of document/file, a natural choice would be to detect the links on a sliding window of sentences in the input stream, that is, to find a connection between claims and evidence separated by at most $n$ other sentences in the stream. This sort of ``locality principle'' especially holds true in such contexts as argumentation mining, where claims and supporting evidence are typically near to each other in the input text~\cite{Eger2017}. 
\end{itemize}

The implementation of the second phase of \ac{MARGOT} for stream processing, presented in Listing~\ref{lst:2nd-phase-stream} enables our system to accommodate both semantics. 
We do not report the details of the first phase, since it is rather similar to that of batch processing.

\Suppressnumber
\begin{lstlisting}[language=Scala,label={lst:2nd-phase-stream},caption={Distributed implementation of \ac{MARGOT}'s second phase (streaming mode).},
float=tp, floatplacement=tbp,
columns=flexible,breaklines=true,breakatwhitespace=true]
Input: claims, evidence /*sentences containing claims or evidences*/, linkModel /*SVM classification model*/, mode /*file-wise or window-wise*/, winDim /*dimension of the join window for window-wise link detection*/
Output: links /* (claim,evidence) pairs containing an argumentative link*/ 
|\Reactivatenumber|
if (mode==file-wise){
  //join all claims and evidences in each file 
  claimsCollection = claims
    .updateStateByKey(updateClaimCollection)  //emits a stateful collection of all the claims encountered since the beginning of the streamed document |\label{lst:-phase-stream:stateful} |
  pairs=claimsCollection.join(evidence)  |\label{lst:-phase-stream:noWinJoin} |
}else{
  //join on a specific window dimension |\label{lst:-phase-stream:winStart} |
  pairs = claims
    .window(winDim)
    .join(evidences.window(winDim)) |\label{lst:-phase-stream:winEnd} |
}
links = pairs
  .mapPartitions((partition) => {
    svmL = SvmClassify.load(linkModel)
    outList : List[(String, (String, String),Double)] = List()
    l = partition.toList 
    l.foreach( x => { //compute link score of each sentence in partition
      filename = x._1
      claim = x._2._1._1
      evid = x._2._2._1
      fvC = x._2._1._2
      fvE = x._2._2._2
      linkScore = svmC.getScore(fvC, fvE)
      outList = (filename, (claim, evid, linkScore)) :: outList  })
  outList.iterator })
 .filter(_._2._3>0) //only pairs with linkScore > 0
\end{lstlisting}

To perform link detection on a sliding window of the input stream (lines~\ref{lst:-phase-stream:winStart} to~\ref{lst:-phase-stream:winEnd}), the flows of claim and evidence collections are sliced using a basic \coding{window} operation before performing the join. In this way, only the claims and evidence inside each window are paired.
To perform link detection with scope file, instead, a stateful operation is performed to maintain a growing collection of claims encountered in each streamed file (line~\ref{lst:-phase-stream:stateful}). This set of past claims is joined with each new evidence detected in the stream (line~\ref{lst:-phase-stream:noWinJoin}).

\section{Empirical Analysis}\label{sec:experiments}
The objective of the analysis we are going to discuss here is gaining a quantitative understanding of the performance enhancement that can be obtained when a \ac{ML} task is distributed on a network of computing nodes. We are not interested here in evaluating the accuracy of the \ac{ML} methods themselves, since that measure would be independent of the  architecture being parallel or otherwise, and anyway it has been  studied in  previous works \cite{DBLP:journals/eswa/0001T16,DBLP:conf/ijcai/LippiT15}.
Indeed, \ac{MARGOT} here works as a case study for investigating the scalability of a \ac{MLaaS} application, in case of both batch and stream processing.

\subsection{Simulation setup}
\label{subsec:setup}

We evaluate the performance of the proposed distributed system for argumentation mining on a cluster of 126 physical nodes. One of these machines, configured as a Spark master, coordinates the work of the others 125 slaves. All the computers are equipped with 8 CPUs, 16GB of RAM, and a 400GB hard disk. The nodes are interconnected by a 100Mbit/s bandwidth local network.
We evaluate the distributed version of \ac{MARGOT} on a collection of input files downloaded from the Project Gutenberg web site.\footnote{\url{https://www.gutenberg.org}} As we need to operate on texts containing a significant number of claims and evidence, common novels are not suitable as input. We therefore restrict our attention to essays in English language. The considered dataset includes 50 files, yielding 466,483 total sentences.
The complete source code of the distributed version of \ac{MARGOT} is available on GitHub \cite{pm}.

\subsection{Evaluation approach}
\label{subsec:approach}
We run separate experiments to evaluate the performance of the system in the two execution modes: with batch input documents and text streams.

To investigate batch processing,  we stored the documents downloaded from Project Gutenberg into a \ac{HDFS} \cite{hdfs}, which automatically slices and distributes the files on the network of computing nodes.
When studying the system for stream processing, instead, we must consider an input flow of text with a certain rate, measured in bytes per second.
In that case, it is crucial for the system to be able to perform all the steps in the pipeline, while keeping up with the input rate. If the computation is slower than the input flow, not only will the system introduce an increasing delay in the time to emit the output, but also the buffer area employed by Spark to temporarily store the data waiting to be analyzed may eventually become saturated. 
The Spark Streaming module treats the flow by periodically slicing it into portions called micro-batches, which are later distributed on the network and separately processed on each node. The period of micro-batch slicing is a configurable parameter. As a general recommendation large micro-batch period helps to keep up with high input rates at the price of an increased latency in the results. Because our goal is not to evaluate the performance of the Spark Streaming's micro-batch processing mechanism, but to evaluate the scalability of the system, we fixed the batch time to 100 seconds for all the tests on streams.

Real-world stream processing services usually experience increases and decreases of the input rate over the day. In order to evaluate the proposed \ac{MLaaS} application we progressively increase the input rate during each test, so as to identify the maximum input rate that the system can sustain before it starts falling behind.  

In both streaming and batch cases, we conduct three scalability tests:
\begin{itemize}
    \item \emph{Test 1} -- experiment with increasing input size (i.e., increasing file dimension for batches, and larger window for streams). The objective is to determine the scalability of the overall \ac{MLaaS} application.
    \item \emph{Test 2} -- experiment with increasing number of \emph{(key, value)} pairs emitted by the first phase (i.e., the number of emitted claims and evidence that would be later joined and processed in the second phase). The objective is to study the effect of the aggregation step bottleneck on the performance.
    \item \emph{Test 3} -- experiment with increasing number of support vectors in the employed \ac{SVM} model. The objective is to study the impact of computationally demanding \ac{ML} tasks.\footnote{Because the features of each sentence must be compared with all the vectors, it is well-known that for an \ac{SVM} classifier larger models yield longer computation times.}
\end{itemize}
We shall remark that \emph{Tests 1} and \emph{2} are independent of the specific \ac{ML} application considered, whereas \emph{Test 3} has been specifically conceived in the context of \ac{MARGOT}, because it is strictly related to the kind of operations conducted in its pipeline. However, it is of general interest, since \ac{SVM} classifiers are widely popular due to their excellent performance in a large variety of tasks.

\subsection{Results}
\label{subsec:results}
\begin{figure*}[b!]
\centering
\subfloat[]{\includegraphics[width=.355\textheight]{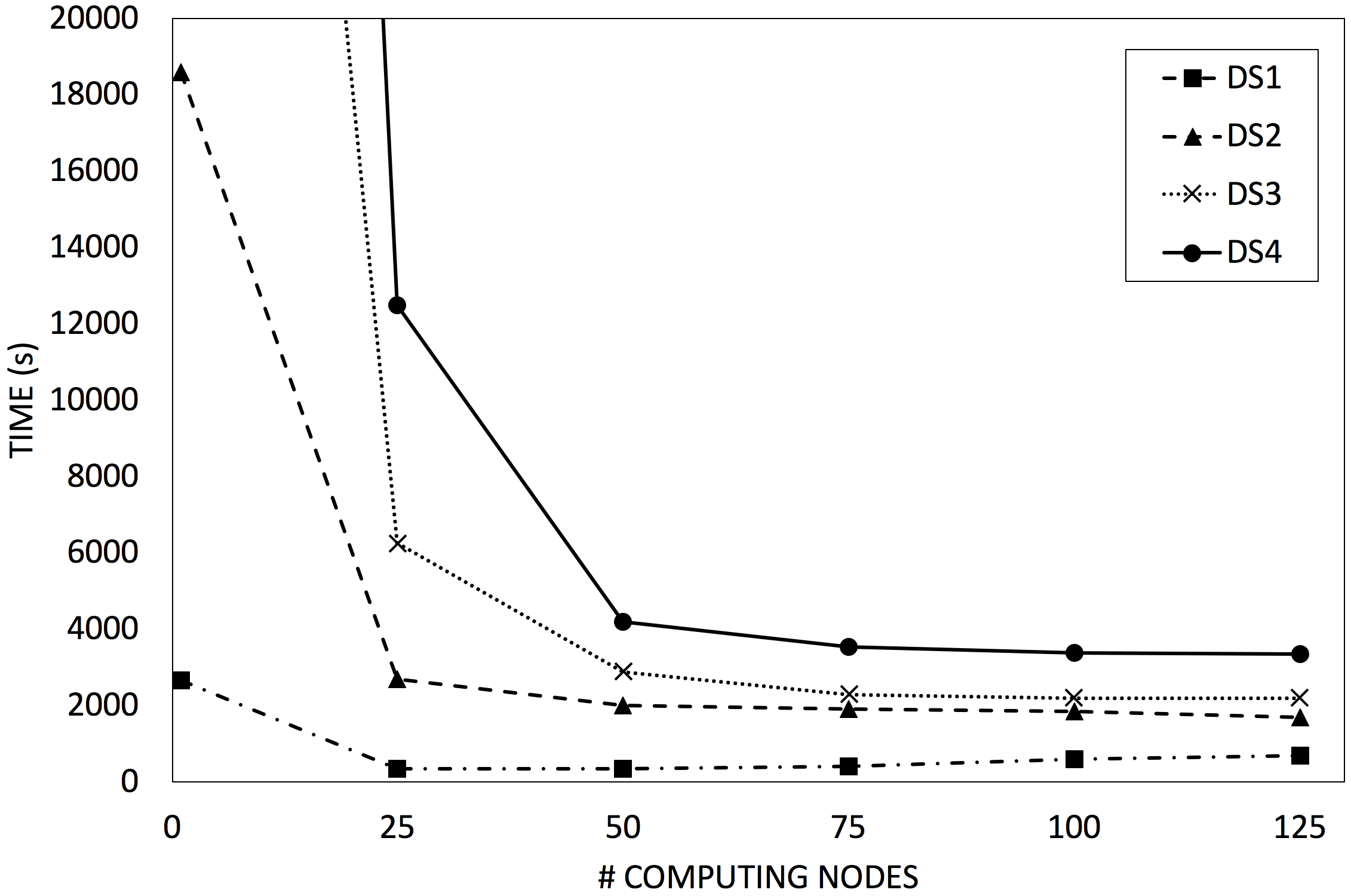}
\label{fig:b-data}}
\subfloat[]{\includegraphics[width=.35\textheight]{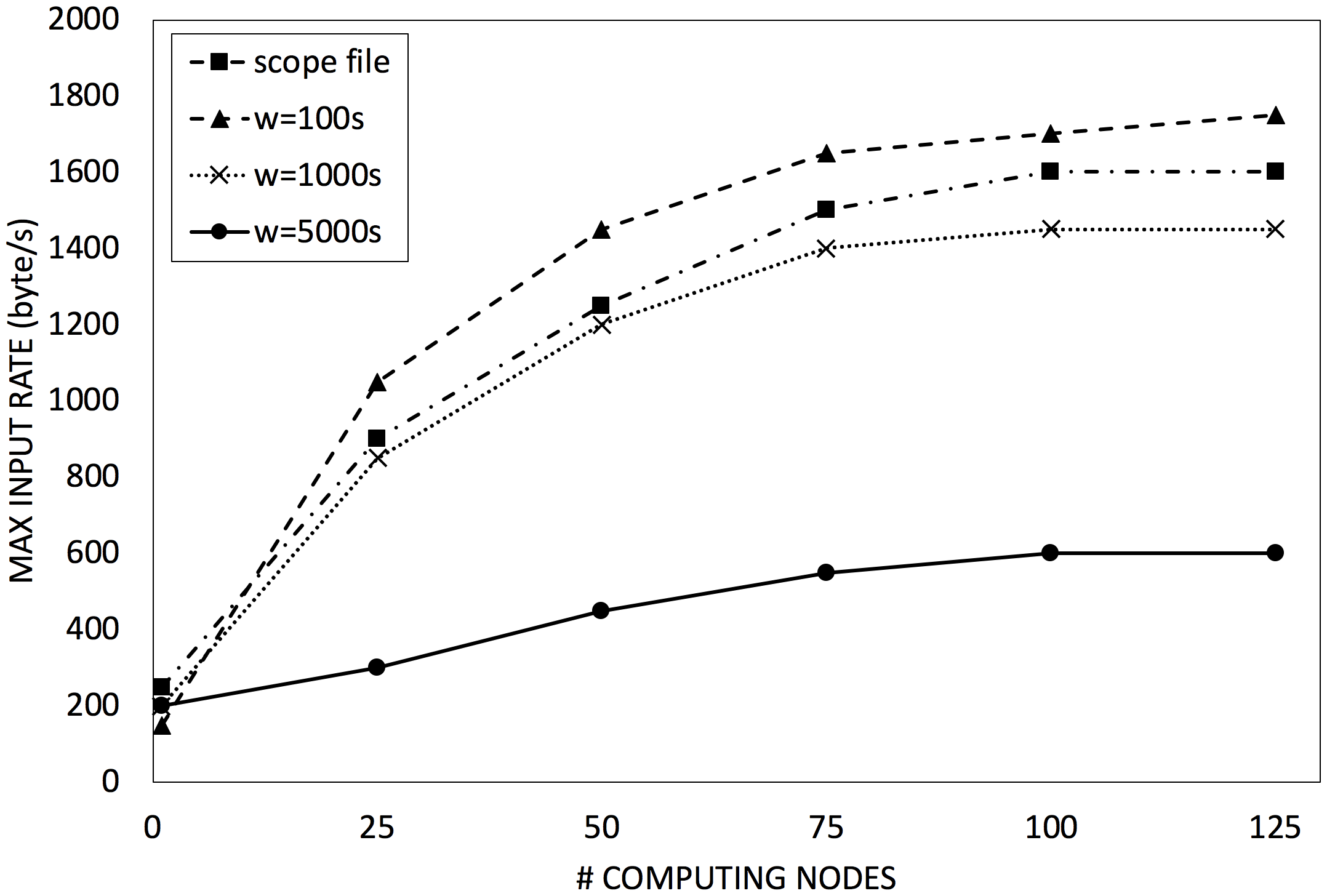}
\label{fig:s-data}}
\caption{System performance by varying size/rate of the input, and computing nodes. Batch mode (left) and stream mode (right).}
\label{fig:data}
\end{figure*}

Concerning \emph{Test 1}, Figure~\ref{fig:data} illustrates the scalability of the distributed system by increasing amounts of input data.
In particular, the plot on the left (Figure~\ref{fig:b-data}) shows the time required to process datasets of different size using batch processing with  increasing numbers of computing nodes. The size of datasets are reported in Table~\ref{tab:ds}.

\begin{table}
\centering
 \begin{tabular}{||c|c|c|c|c||} 
 \hline
 Dataset & Sentences & Claims & Evidence & Links \\ [0.5ex] 
 \hline\hline
 DS1 & 9,783 & 1,244 & 2,739 & 4,489 \\ 
 \hline
 DS2 & 67,917 & 8,050 & 16,267 & 9,827 \\ 
 \hline
 DS3 & 233,254 & 13,597 & 76,173 & 29,027 \\ 
 \hline
 DS4 & 466,483  & 27,193 & 152,345 & 58,464 \\ 
 \hline
\end{tabular}
\caption{Details of the datasets employed in batch processing evaluation of Figure~\ref{fig:b-data} }
\label{tab:ds}
\end{table}

As desired, the total execution time greatly benefits from the introduction of additional nodes. The most significant improvements are observed between 1 and 50 nodes. After 50 nodes, the cost of distributing the tasks on the network balances off the benefits yielded by the additional computational resources. Furthermore, when dealing with small input files such as the ``DS1'' series, a slight performance loss is observed as the number of nodes increases from 25 to 50 and above. That could be the effect of the overhead generated by partitioning and distributing in the network small amounts of data.

The other plot (Figure~\ref{fig:s-data}) illustrates the system's performance with text streams. The y-axis reports the maximum input rate in bytes per second that the system can tolerate without falling behind. 
The graph has been plotted by periodically increasing the input rate and checking when the processing time of each micro-batch in the stream started to exceed the configured micro-batch period. Hence, the higher is the curve in the figure, the better is the system performance. 
We have made several experiments by varying the windows size. For example, with $w = 5,000 s$ we indicate the performance of the system when claims and evidence are joined over a window of 5,000 seconds, that is fifty times bigger than the micro-batch period. In this case, when employing 125 nodes, the system cannot keep up with an input frequency higher than 600 bytes/s. If we assume the average sentence length to be 200 characters, this means that the system cannot process more than 3 sentences per second. Although such a performance may seem unimpressive at a first glance, we should consider the sheer number of claim/evidence pairs to be analyzed by the link classifier. In particular, with a 600 bytes/s input rate and a 100s micro-batch period, a 5,000 seconds window contains around 15,000 sentences.  

\begin{figure*}[b!]
\centering
\subfloat[]{\includegraphics[width=.356\textheight]{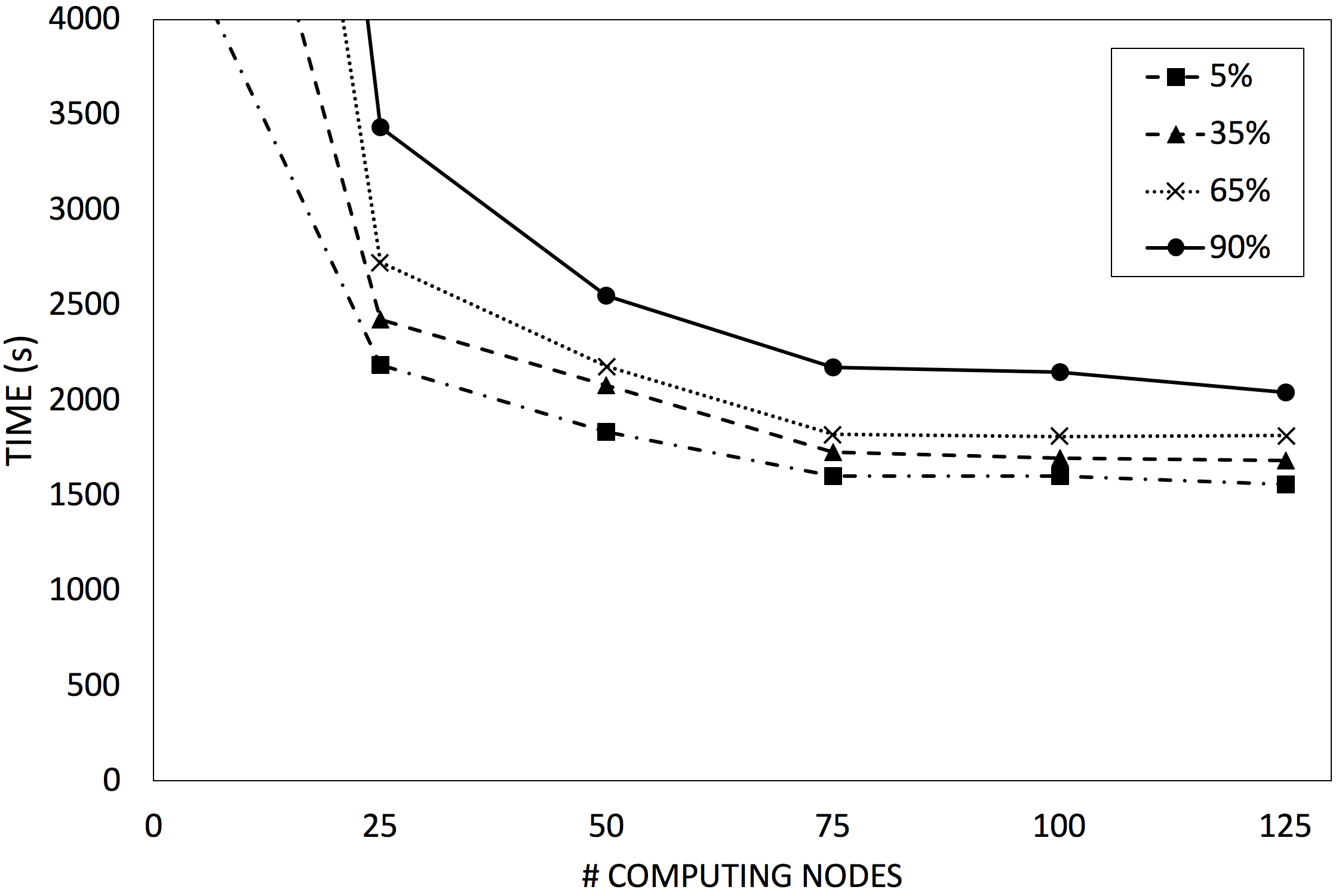}
\label{fig:b-ecThr}}
\subfloat[]{\includegraphics[width=.35\textheight]{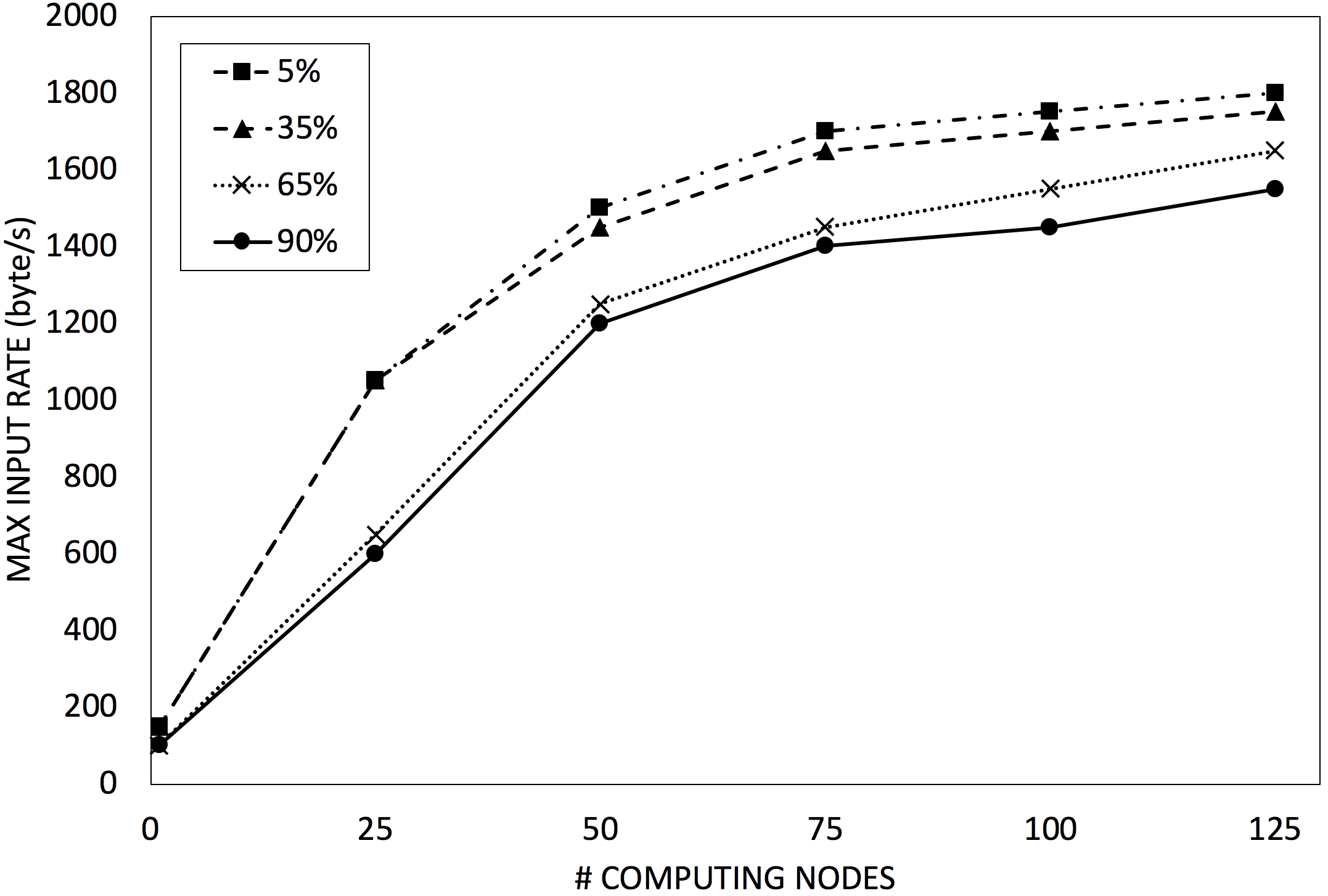}
\label{fig:s-ecThr}}
\caption{Scalability of the system by increasing amount of \emph{(key, value)} pairs emitted by the first phase. Left: batch processing; right: stream processing.}
\label{fig:ecThr}
\end{figure*}

Figure~\ref{fig:s-data} also reports the performance in the ``scope-file'' series, when no window is employed, but instead the claim/evidence join is executed on a per-file basis. The performance are worse than those obtained with a window of size 100s (up to around 900 sentences per window processed on 125 nodes) and better than 1,000s (around 72,000 sentences per window). As the average number of sentences in each file is 9,000, the position of the ``scope-file'' curve between series ``w=100s'' and ``w=1,000s'' appears reasonable.

Figure~\ref{fig:ecThr} illustrates the results of \emph{Test 2}.
In order to obtain different amounts of elements to be analyzed in the second phase, in Test 2 we artificially varied the filtering thresholds that \ac{MARGOT} uses to identify claims and evidence at the end of the first phase. The names of the series (5\%, 35\%, 65\% and 90\%) report the percentage of input sentences that reach the second phase. As expected, both batch (Figure~\ref{fig:b-ecThr}) and stream (Figure~\ref{fig:s-ecThr}) processing reveal a significant effect of this parameter on the processing time and the maximum input rate. This confirms the bottleneck effect of the aggregation step in the \ac{ML} pipeline. Nonetheless, the scalability trend is evidently maintained for all the series in the graphs.

\begin{figure*}[t!]
\centering
\subfloat[]{\includegraphics[width=.355\textheight]{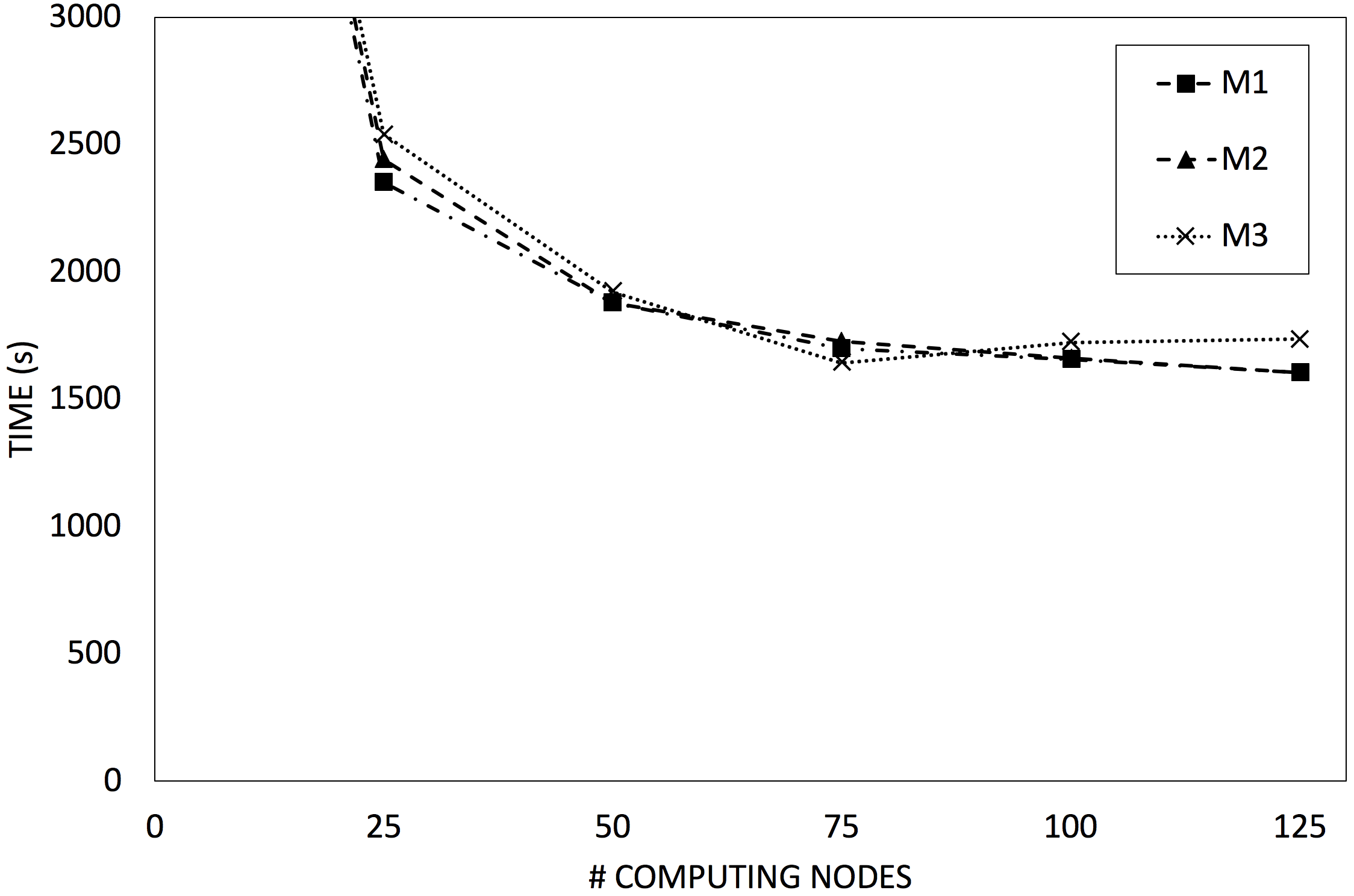}
\label{fig:b-model}}
\subfloat[]{\includegraphics[width=.35\textheight]{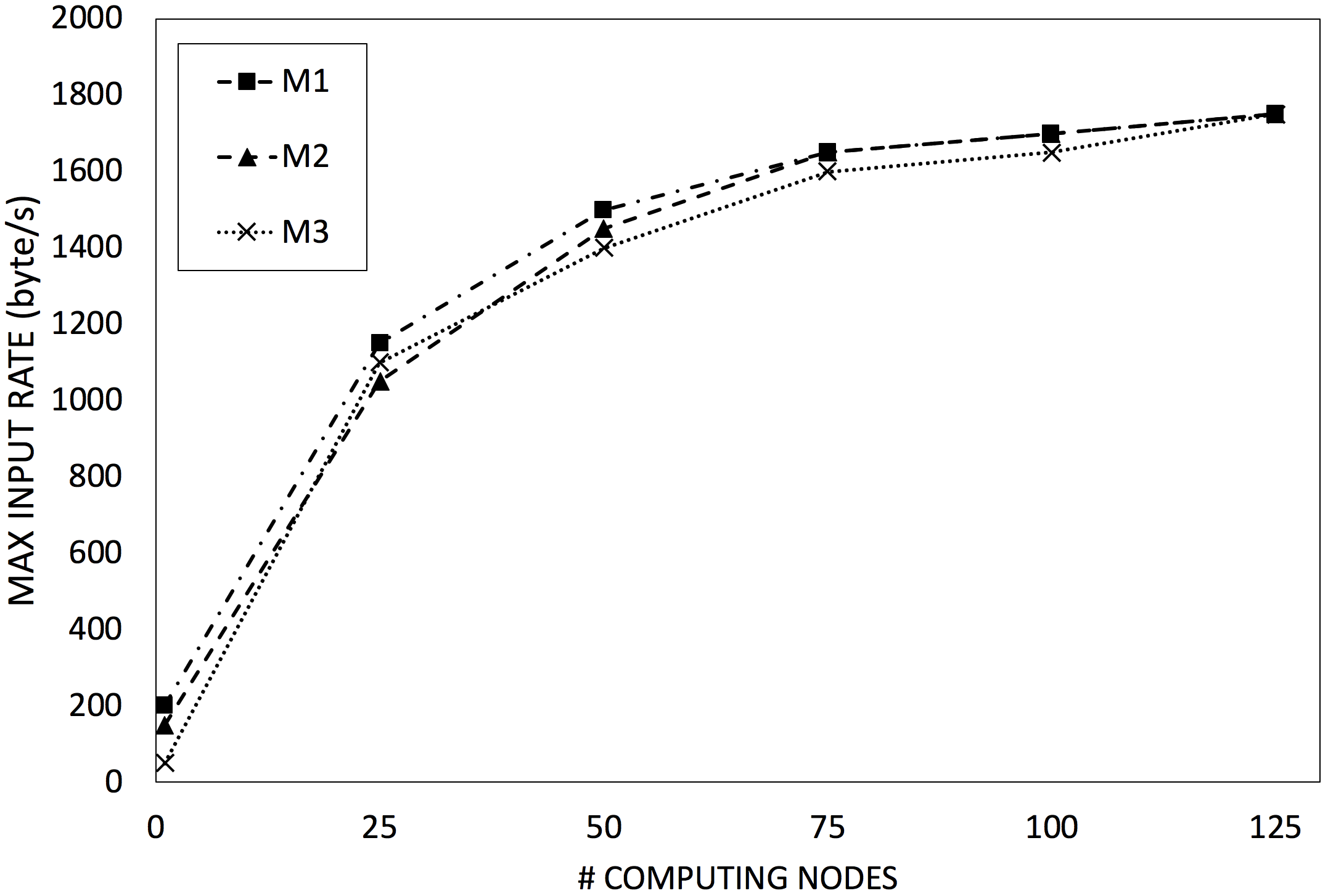}
\label{fig:s-model}}
\caption{System performance by varying complexity of the model. Left: batch processing; right: stream processing.}
\label{fig:model}
\end{figure*} 

\begin{table}
\centering
 \begin{tabular}{||c|c|c|c|c||} 
 \hline
 Model & Dimension (MB) & Support Vectors \\ [0.5ex] 
 \hline\hline
M1 & 6.7 & 7,085 \\ 
 \hline
M2 & 17.3 & 18,604 \\ 
 \hline
M3 & 29.9 & 30,363\\ 
 \hline
\end{tabular}
\caption{Details of the model employed in \emph{Test 3}}
\label{tab:mod}
\end{table}

Figure~\ref{fig:model} illustrates the results of \emph{Test 3}, whose aim was to study the impact of the number of support vectors in the link prediction model on the system's performance. Table~\ref{tab:mod}  summarizes the details of each model. 

As the \ac{SVM} classifier checks the features of each (\emph{claim,evidence}) pair with all the support vectors in the model, one could imagine that larger models cause longer computation times. Instead, rather surprisingly, we observed that the effect of this parameter on the overall batch (Figure~\ref{fig:b-model}) and stream processing (Figure~\ref{fig:s-model}) performance is insignificant and that scalability is not visibly affected.

\section{Discussion}\label{sec:discuss}

Our empirical analysis indicates a promising scalable behaviour in all the considered scenarios. In addition, our study was instrumental in identifying challenges and opportunities that could arise in general from the parallelization of other \ac{ML} services.

\subsection{Challenges of \ac{ML} pipeline parallelization}
Independently of the particular \ac{ML} application, the implementation of the pipeline initially depicted in Figure~\ref{fig:g-pipeline} on a network of cooperating computing nodes presents three major architectural and technological challenges. First of all, while the first phase enjoys the benefits of a highly parallelizable structure, the aggregation step in the second phase needs careful consideration, to prevent it from turning into a bottleneck. Indeed, combining several records together is an expensive operation in distributed environments because it causes massive data shuffle over the network. This is actually a known issue for many MapReduce applications. In case of \ac{MLaaS}, it may be crucial to apply, if possible, a filtering operation aimed to reduce the input space for the aggregation function. For instance, one could anticipate some steps that conceptually may belong further down in the pipeline, in an effort to narrow down the data emitted at the end of the first phase.

A second challenge has to do with the trained model supervised or semi-supervised \ac{ML} applications commonly use in order to process the input data and provide predictions.
When a distributed engine is employed, each computing node must be provided a copy of such a model alongside with the input data. 
However, depending on the application and the underlying \ac{ML} technologies, the dimension of the trained model can be significant. 
Nonetheless, if the learning process is not continuous, meaning that the model evolution is limited to the training phase, such a model is stable during the whole prediction process. Accordingly, 
the model could be distributed to all computing nodes at the beginning of the computation once and for all (e.g., thanks to mechanisms such as the broadcast variables of Apache Spark), and then employed by each node independently to process/filter its portion of the input.

A final technological challenge has to do with the third-party software  \ac{ML} applications often employ at different stages of the pipeline. The integration of third-party methods into the framework of a distributed data processing engine might not be a straightforward operation. Ad-hoc solutions (such as Spark's \textit{mapPartitions}) might be required, for instance, to limit the number of calls to external processes executing the third-party software. 

\subsection{Other aspects of \ac{MLaaS} on large-scale data processing engines}

The implementation of \ac{MLaaS} on large-scale data processing engines not only offers architectural and technological challenges, but it also often requires making some detailed choices whose impact on the performance of the offered service may be crucial. 
For example, although the aggregation step of the second phase works by collecting together the output of the previous processing, performing the subsequent processing/filtering step on a single node is not mandatory, and indeed it is generally worth avoiding. If a certain degree of parallelism is possible (because for example the instances emitted by the aggregation step can be considered independently for one another, as was for MARGOT's  claim/evidence pairs), a good practice would be to split the second phase too into tasks that can be carried on concurrently. Modern big data analytics engines support the implementation of distributed programs not strictly limited to a map and a reduce phase. More complex variants and combinations of the two are possible. As far as \ac{MLaaS} is concerned, a further map step following the aggregation could contribute to boosting the performance of the second phase.

Furthermore, when the processing/filtering steps involve a large trained model, the task of loading such model can be time consuming and should be performed carefully. Consider for example a natural language processing system whose first phase classifies the sentences of a text according to the features of a large trained model. 
In such a system it is certainly not recommended to load the model for each sentence. A much preferable solution would be to load the model once, and then use it to analyze a consistent number of phrases together. 
Nonetheless, grouping too many sentences together reduces the degree of parallelism of the pipeline step, thus slowing down the computation. Depending on the specific predictive task to be carried out and on the degree of parallelism allowed by the underlying infrastructure, a trade-off must be found between the need to avoid unnecessarily repeating costly operations (such as model loading) and that of splitting the work to speed up the computation.

A similar conundrum regards the features extracted from the input data which might be used in multiple subsequent steps of the pipeline. Feature extraction may be a costly operation, in which case it would be worthwhile performing it only once, and then hand over the features to the following steps until they are no longer needed. This is indeed the solution adopted by \ac{MARGOT}, where the features extracted in the first phase are passed along the pipeline together with the relevant sentence. However, one must consider that between each pair of steps of the pipeline a data shuffle over the network might occur. Therefore, if feature extraction produces large outputs, the shuffling of such data on a network with limited bandwidth might cause poor performance. In these cases, there is a trade-off to consider, between recomputing large features when needed and shuffling them between nodes.

Finally, a common feature of \ac{MLaaS} applications is the requirement to accommodate a variety of input modes, in particular document batches and data streams. From a developer's standpoint, a shift of perspective is unavoidable when passing from a \ac{ML} application working on input files that are already materialized and stored in a specific location (i.e., batch processing) to the analysis of a flow of input data (i.e, stream processing). Nonetheless, some relatively recent MapReduce-oriented platforms \cite{flink,spark} allow this shift at the price of slight changes in the application implementation. 
Since the first phase of our reference \ac{ML} pipeline operates in theory on each input instance independently, it could operate on batches and streams alike. 
The second phase instead is likely to be dependent on the input mode, because it focuses on the aggregation and processing of the previous step's output. Indeed, for stream processing aggregation entails the need to specify not only the collections of data to be merged, but also the period over which such operation must be performed. For example, aggregation could be based on the last $n$ occurrences in the flow, on a specific time window, or on all the data received so far. Each option would have a different meaning from the others and may produce completely different results. In the case of batch processing, instead, the aggregation task has a less varied range of semantics, because there are no concepts like ``time window'' or ``arrival instant'' of an input entry. 

\section{Conclusions}\label{sec:conclusions}
A growing number of ML applications are being deployed as ready-to-use, already trained services for the end-users. This calls for implementing distributed architectures able to scale up such services to broader user communities, and larger data collections.

In this paper, we presented a distributed architecture inspired by the MapReduce paradigm, which could be used to parallelize the prediction phase of a typical \ac{ML} pipeline. We conducted experimental results on a real-world text mining application case study. We also discussed how the methodology is general enough to be applied to many other different scenarios. We considered both batch and stream processing, and studied the performance gain that can be achieved by this architecture under many angles.

An interesting open challenge is how to effectively extend this architecture to accommodate  \ac{ML} applications dealing with structured data, such as sequences, trees, and graphs. In that case, since the relations between the input data are as relevant as the data themselves, it is unlikely that the first phase could be implemented through a simple split operation, followed by independent map processes. Instead, we expect that a more elaborated slicing procedure would be needed at the beginning of the pipeline, in order to correctly partition the data across the nodes. 
A technique used to achieve a desired level of parallelism with input sequences of data in another application domains involved data replication~\cite{DBLP:journals/fgcs/LoretiCCM18}. The idea was to divide sequences into slices and replicating the data on the extremities, before assigning each slice to a different nodes. The identification of patterns for the split step that can be applied and reused in case of more elaborated input structures would be an interesting subject for future investigation.
An even more challenging setting would be the case where the input data has a highly connected structure, hindering any kind of split and re-partition of the work. Then a completely different approach could be explored: instead of slicing the data, one could provide each node with the whole dataset, but only a portion of the trained model. In this way, each machine of the distributed architecture could conduct a lightweight prediction analysis on all the input. Like in the pipeline described in this work, the results of those analyses will have to be conveniently aggregated in the following phase.

\section*{Acknowledgements}{This work has been partially supported by the H2020 Project AI4EU (g.a. 825619).}





\section*{References}








\begin{acronym}
\acro{HDFS}{Hadoop Distributed File System}
\acro{MARGOT}{Mining ARGuments frOm Text}
\acro{ML}{Machine Learning}
\acro{MLaaS}{Machine Learning as a Service}
\acro{SVM}{Support Vector Machine}
\acrodefplural{SVM}{Support Vector Machines}
\acro{SSTK}{SubSet Tree Kernel}
  
\end{acronym}
\end{document}